\documentclass[12pt]{article}

\usepackage{amsmath,amsfonts}

\begin{document} 

\begin{titlepage}

\hrule 
\leftline{}
\leftline{Preprint
          \hfill   \hbox{\bf CHIBA-EP-110}}
\leftline{\hfill   \hbox{October 1998}}
\vskip 5pt
\hrule 
\vskip 1.0cm

\centerline{\large\bf 
Self-Duality of Super D3-brane Action
on $AdS_{5}\times S^{5}$ Background 
}

\vskip 1cm

\centerline{{\bf 
Tadahiko Kimura$^{1}{}^{\dagger}$
}}  
\begin{description}
\item[]{\it  
$^1$ Department of Physics, Faculty of Science,
  Chiba University, Chiba 263, Japan
  }
\item[]{$^\dagger$ 
  E-mail:  kimura@cuphd.nd.chiba-u.ac.jp 
  }
\end{description}

\centerline{{\bf Abstract}} \vskip .5cm
We show that the super D3-brane action on $AdS_5 \times S^5$
background recently constructed by Metsaev and Tseytlin is
exactly invariant  under the combination of the 
electric-magnetic duality transformation of the worldvolume gauge
field and the $SO(2)$ rotaion of $N=2$ spinor coordinates. 
The action is shown to satisfy the Gaillard-Zumino duality
condition, which is a necessary and sufficient condition for a
action to be self-dual. Our proof needs no gauge fixing for
the $\kappa$-symmetry.



\vskip 0.2cm  


\end{titlepage}

\pagenumbering{arabic}

\newpage
\section{Introduction}
\setcounter{equation}{0}

Recently much attention is concentrated on the connection between
string theory and M-theory on $AdS_{p+2}\times S^{D-p-2}$ and
extended superconformal theories in $p+1$ spacetime dimensions 
\cite{M, GKP,W}.
$AdS_5\times S^5$ describes a maximally supersymmetric vacuum
(besides flat space) of type IIB supergravity and is the 
near horizon geometry of the D3-brane solution.
$AdS_4 \times S^7$ and $AdS_7 \times S^4$ are maximally
supersymmetric vacua of 11-dimensional supergravity and are the
near horizon geometries of the M2- and M5-branes, respectively.

The actions of super $p$-branes in these near horizon
background are described by a Dirac-Born-Infeld (DBI)type term
and a Wess-Zumino term, and are invariant under local
diffeomorphisms of the worldvolume, the $\kappa$-symmetry and the
global isometries of the background.

Recently the super p-brane actions on 
$AdS_{p+2}\times S^{D-p-2}$ background geometries have been 
constructed for the superstring \cite{MT1, KRR, KR, P, KT}
and the D3-brane \cite{MT2} on the  $AdS_5\times S^5$ background
, for the M2-brane
on the $AdS_4 \times S^7$ solution \cite{DFFFTT, dPPS} and M5-brane
on the $AdS_7 \times S^4$ solution \cite{C}.

On the other hand electric-magnetic duality relations among 
various super p-brane effective actions on the flat background have
been found in semiclassical approximation \cite{T, APPS} and for
some cases on quantum level\cite{IIKK,O1,IIK1}. It will be of great
interest to investigate whether these duality  relations hold valid
for  super $p$-brane actions on $AdS$  background. 
In some recent literatures the quantum equivalence of super 
D-string and F-string on $AdS_5\times S^5$  and the semiclassical
self-duality of super D3-brane on $AdS_5\times S^5$ have already 
been discussed \cite{O2,O3}. However in these works proofs of
the self-duality relied on a specific gauge choice for the $\kappa$
-symmetry.

The purpose of the present paper is to show that the super D3-brane
action on $AdS_5 \times S^5$ constructed by Metsaev and Tseytlin 
\cite{MT2} is exactly self-dual. The action is shown to satisfy the
Gaillard-Zumino(GZ)  duality condition \cite{GZ1, GZ2, GR1}. 
Igarashi et.at. have shown that the GZ-condition is actually
necessary  and sufficient condition for the theory to be self-dual
\cite{IIK2} and they have shown that the super D3-brane action  on
the flat background indeed satifies the GZ-duality condition
\cite{IIK1}. 

This paper is organized as follows. In the next section a brief
review of the super D3-brane action on $AdS_5 \times S^5$
constructed  by Metsaev and Tseytlin by the coset formalism
\cite{MT1,MT2} is given.  In section 3 the
super D3-brane action on $AdS_5 \times S^5$ is shown to satisfy the
GZ duality condition, thereby establishing its exact self-duality
without  resort to any semiclassical approximation.

\section{Super D3-brane Action on $AdS_5 \times S^5$}
\setcounter{equation}{0}
In this section, for later use, we briefly review the D3-brane
action constructed by Metsaev and Tseytlin \cite{MT2}. 

Supervielbein superfields on the $AdS_5 \times S^5$ background are
explicitly constructed by the coset formalism \cite{MT1,KRR}. 
In the explicit coset parametrization $G(X^{\hat{m}},\theta)=
g(X^{\hat{m}})e^{\theta Q}$, where $X^{M}=(X^{\hat{m}},\theta)$ and
$Q$ are $N=2$ superspace coordinate and supercharge, respectively,
the supervielbein $L^A=(L^{\hat{a}}, L)$ and superconnection
$L^{\hat{a}\hat{b}}$ are explicitly given as follows;

\begin{eqnarray} 
L^{\hat{a}}=e^{\hat{a}}-2i\bar{\theta}
\Gamma^{\hat{a}}W(\theta)D\theta, L=V(\theta)D\theta,
\end{eqnarray}

\begin{eqnarray}
L^{ab}=\omega^{ab}+2\bar{\theta}i\tau^{(2)}
\Gamma^{ab}\sigma_-W(\theta)D\theta,
L^{a^\prime b^\prime}=\omega^{a^\prime b^\prime}
+2\bar{\theta}i\tau^{(2)}
\Gamma^{a^\prime b^\prime}\sigma_-W(\theta)
D\theta,
\end{eqnarray}
where $a,b$ ($a^{\prime}, b^{\prime})$ are $AdS_5$ ($S^5$)
coordinate indices and run 0,1,2,3,4 (5,6,7,8,9), and
$D\theta$,$V(\theta)$ and
$W(\theta)$ are defined by
\begin{eqnarray}
D\theta =(d +
\frac{1}{4}\omega^{\hat{a}\hat{b}}\Gamma^{\hat{a}\hat{b}}
+\frac{1}{2}\tau^{(2)}e^{\hat{a}}\sigma_+\Gamma^{\hat{a}})\theta.
\end{eqnarray}
\begin{eqnarray}
V(\theta)=\frac{\sinh\sqrt{m(\theta)}}{\sqrt{m(\theta)}}
= 1 + \frac{1}{3!}m(\theta) + \frac{1}{5!}m(\theta)^2 + \cdots ,
\\
W(\theta)=\frac{\cosh\sqrt{m(\theta)}-1}{m(\theta)}
= \frac{1}{2!} + \frac{1}{4!}m(\theta) + \frac{1}{6!}m(\theta)^2
+ \cdots,
\end{eqnarray}
\begin{eqnarray}
m(\theta)=-\sigma_+\Gamma^{\hat{a}}i\tau^{(2)}\theta\bar{\theta}
\Gamma^{\hat{a}}
+\frac{1}{2}\Gamma^{ab}\theta\bar{\theta}i\tau^{(2)}
\Gamma^{ab}\sigma_-
-\frac{1}{2}\Gamma^{a^\prime b^\prime}\theta\bar{\theta}
i\tau^{(2)}\Gamma^{a^\prime b^\prime}\sigma_-.
\end{eqnarray}
In the above equations Pauli matrices $\tau^{(i)}$ operate
on internal $N=2$ space and Pauli matrices
$\sigma_i$ stand for $32\times32$ matrices $1\times 1\times 
\sigma_i$.

The D3-brane action depends on the coset superspace
coordinates
$X^{M}=(X^{\hat{m}},\theta)$ and U(1) world-volume gauge field
strength
$F=dA=\frac{1}{2}F_{ij}d\sigma^i\wedge d\sigma^j,
F_{ij}=\partial_iA_j-\partial_jA_i$, where $i$ and $j$ run over the
worldvolume indices 0,1,2,3. The reparametrization and
$\kappa$-symmetry invariant super D3-brane action in the $AdS_5
\times S^5$ background consists of two terms; i.e. the
Dirac-Born-Infeld(DBI) term and the Wess-Zumino term:
\begin{eqnarray}
S=S_{DBI}+S_{WZ}.
\end{eqnarray}

The DBI term is given by
\begin{eqnarray}
S_{DBI}=-\int_{M_4}d^4\sigma\sqrt{-det(G_{ij}+{\cal F}_{ij})},
\end{eqnarray}
where $G_{ij}$ is the supersymmetric induced worldvolume metric
defined by
\begin{eqnarray}
G_{ij}=L_i^{\hat{a}} L_i^{\hat{a}}
= \partial_i X^M L_M^{\hat{a}}\partial_j X^N L_N^{\hat{a}},
L^{\hat{a}}=d\sigma^i L_i^{\hat{a}},
\end{eqnarray}
and two-form superfield ${\cal F}$ is the supersymmetric extension
of the U(1) worldvolume gauge field strength defined by
\begin{eqnarray}
{\cal F} = dA + \Omega^{(3)},
\end{eqnarray}
where two-form $\Omega^{(i)}$ is defined by
\begin{eqnarray}
\Omega^{(i)} = 2i\int_0^1 ds\bar{\theta}\hat{L}_s\tau^{(i)}L_s,
\end{eqnarray}
and  $L_s^{\hat{a}} \equiv L^{\hat{a}}(x, s\theta), L_s\equiv
L(x,s\theta)$ and $\hat{L}_s \equiv L_s^{\hat{a}}\Gamma^{\hat{a}}$.

The Wess-Zumino term is given by
\begin{eqnarray}
S_{WZ}=\int_{M_4}\Omega_4=\int_{M_5}H_5, 
\end{eqnarray}
here
\begin{eqnarray}
H_5&=&d\Omega_4 \nonumber \\
&=& i\bar{L}\wedge(\frac{1}{6}\hat{L}\wedge\hat{L}\wedge
\hat{L}i\tau^{(2)} 
+{\cal F}\wedge\hat{L}\tau^{(1)})\wedge L \nonumber \\
&+&\frac{1}{30}(\epsilon^{a_1\cdots
a_5}L^{a_1}\wedge\cdots\wedge L^{a_5}+\epsilon^{a^\prime_1\cdots
a^\prime_5}L^{a^\prime_1}\wedge
\cdots\wedge L^{a^\prime_5}).
\end{eqnarray}
It is convenient to rewrite $\Omega_4$ in the following form;
\begin{eqnarray}
\Omega_4 = C_4 + \Omega^{(1)}\wedge{\cal F},
\end{eqnarray}
then we obtain a differential equation for $C_4$
\begin{eqnarray}
dC_4 +\Omega^{(1)}\wedge d\Omega^{(3)}
&=&i\frac{1}{6}\bar{L}\wedge\hat{L}\wedge\hat{L}\wedge
\hat{L}i\tau^{(2)}\wedge L \nonumber \\
&+&\frac{1}{30}(\epsilon^{a_1\cdots
a_5}L^{a_1}\wedge\cdots\wedge L^{a_5}+\epsilon^{a^\prime_1\cdots
a^\prime_5}L^{a^\prime_1}\wedge
\cdots\wedge L^{a^\prime_5}).
\end{eqnarray}
The equation (2.13) or (2.15) can be explicitly integrated by using
the Maurer-Cartan equations and the $\theta \rightarrow s\theta$
trick and the result is given in \cite{MT2}. However for our purpose
the  explicit form of $C_4$ is not necessary.

As it can be seen from (2$\cdot$1)-(2$\cdot$6), the super
D3-brane action on $AdS_5\times S^5$ is a complicated function of
supercoordinate $\theta$. However to prove the self-duality of
the action (2.7) we need no gauge fixing condition to simplify
the fermionic dependence of the supervielbeins and superconnections.





\section{Self-Duality of Super D3-brane Action\\
on $AdS_5\times S^5$}
\setcounter{equation}{0}

In this section we show that the super D3-brane action on
$AdS_5\times S^5$ reviewed in the last section  satisfies
the Gaillard and Zumino duality-condition, thereby establishing
its exact self-duality without resort to any semiclassical 
approximation.

 First let us recall what is the GZ duality condition. Given
a generic Lagrangian density ${\cal L}(F_{\mu\nu},
g_{\mu\nu},\phi^A) =\sqrt{-g}L(F_{\mu\nu},
g_{\mu\nu},\phi^A)$ in four dimensional spacetime
 which contains a U(1) gauge field strength $F_{\mu\nu}$,
gravitational field $g_{\mu\nu}$ and generic matter fields
$\Phi^A$, the constructive relation is given by
\begin{eqnarray}
\tilde{K}_{\mu\nu} \equiv \frac{\partial L}{\partial
F_{\mu\nu}},
\end{eqnarray}
where the Hodge dual components for the anti-symmetric tensor
$K_{\mu\nu}$ are defined by
\begin{eqnarray}
\tilde{K}_{\mu\nu} \equiv
\frac{1}{2}\eta_{\mu\nu}^{\rho\sigma}K_{\rho\sigma},
\tilde{\tilde{K}}_{\mu\nu}=-K_{\mu\nu},
\end{eqnarray}
where $\eta_{\mu\nu\rho\sigma}
=\sqrt{-g}\epsilon_{\mu\nu\rho\sigma}$,
$\epsilon^{0123}=1$ and the signature of $g_{\mu\nu}$ is (-,+,+,+).

If one defines the infinitesimal $SO(2)$ duality transformation by
\begin{eqnarray}
&&\delta F_{\mu\nu} =\lambda K_{\mu\nu},\delta K_{\mu\nu} = -\lambda
F_{\mu\nu} , \nonumber \\
&&\delta\Phi^A = \xi^A(\Phi),
\end{eqnarray}
then the consistency of the constructive relation (3.1) and the
invariance of the field equations under this $SO(2)$ duality
transformation require the following condition:
\begin{eqnarray}
\frac{\lambda}{4}(F_{\mu\nu}\tilde{F}^{\mu\nu}+
K_{\mu\nu}\tilde{K}^{\mu\nu})
+\delta_{\Phi}{\cal L} =0,
\end{eqnarray}
and the invariance of the energy-momentum tensor requires
$\delta g_{\mu\nu}=0$. We call the condition (3.4) the GZ
self-duality condition \cite{GZ1, GZ2}.

It has been known that the $SO(2)$ duality is lifted to the
$SL(2,R)$ duality by introducing a dilaton $\phi$ and an axion
$\chi$ \cite{GR2}.

Moreover it has also been shown that the GZ condition (3.4) is
actually the necessary and sufficient condition for the action to be
invariant  under the duality transformation (3.3) \cite{IIK2}.
Therefore if one can show an action to satisfy the condition
(3.4) under the transformation (3.3) with suitable transformation
rule for matter fields, then one establishes the exact self-duality
of the theory described by this action without resort to any
semiclassical approximation.

Now let us show that the super D3-brane action on $AdS_5\times  
S^5$ reviewed in the preceding section satisfies the GZ duality
condition under the following $SO(2)$ duality transformation:
\begin{eqnarray}
&&\delta F_{ij} =\lambda K_{ij},\delta K_{ij} = -\lambda
F_{ij} , \nonumber \\
&&\delta\theta = -\lambda\frac{i\tau^{(2)}}{2}\theta, 
\delta\bar{\theta} = \lambda\bar{\theta}\frac{i\tau^{(2)}}{2},
\delta X =0.
\end{eqnarray}
The $N=2$ spinor coodinates transform as an $SO(2)$ doublet.

Let us first examine how various quantities in the action
(2.7) transform under the  above $SO(2)$ duality rotation. The matrix
$m(\theta)$ defined in (2.6) transforms as an adjoint representation
of $SO(2)$, and therefore matrices $V(\theta)$ and $W(\theta)$ obey
the same transformation properties as $m(\theta)$;
\begin{eqnarray}
\delta m(\theta)=-\lambda[\frac{i}{2}\tau^{(2)}, m(\tau)],
\delta V(\theta)=-\lambda[\frac{i}{2}\tau^{(2)}, V(\tau)],
\delta W(\theta)=-\lambda[\frac{i}{2}\tau^{(2)}, W(\tau)].
\end{eqnarray}
From these transformation properties, we can easily show that
the supervielbeins transform as follows;
\begin{eqnarray}
\delta L^{\hat{a}}=0, 
\delta L = -\lambda\frac{i\tau^{(2)}}{2}L,
\delta\bar{L} &=& \lambda\bar{L}\frac{i\tau^{(2)}}{2},
\end{eqnarray}
namely the $N=2$ spinor components of the supervielbeins  $L$ and
$\bar{L}$ transform as $SO(2)$ doublets. 

From (3.5) and (3.7), we find  $\Omega^{(1,3)} =2i\int_0^1
ds\bar{\theta}\hat{L}_s\tau^{(1,3)}L_s$ defined in (2.11) transform
as an $SO(2)$ doublet and $\Omega^{(2)}$ is an $SO(2)$ singlet:
\begin{eqnarray}
\delta\Omega^{(1)}= \lambda\Omega^{(3)},
\delta\Omega^{(3)}=-\lambda\Omega^{(1)},
\delta\Omega^{(2)}= 0.
\end{eqnarray}  

From (3.8) and (2.15) the transformation of $C_4$
is easily obtained. Noticing  the right hand-side of (2.15) is
invariant under
$SO(2)$ rotation, the $SO(2)$ transformation of (2.15) yields
\begin{eqnarray}
d\delta C_4 = -\frac{\lambda}{2}d(\Omega^{(1)}\wedge\Omega^{(1)}+
\Omega^{(3)}\wedge\Omega^{(3)}).
\end{eqnarray}
Then we obtain the transformation rule for $C_4$ up to exact forms;
\begin{eqnarray}
\delta C_4 = -\frac{\lambda}{2}(\Omega^{(1)}\wedge\Omega^{(1)}+
\Omega^{(3)}\wedge\Omega^{(3)})
\end{eqnarray}

Now we are ready to prove the duality symmetry of the super
D3-brane action on $AdS_5\times S^5$. First let us calculate
$\frac{\lambda}{4}(F_{ij}\tilde{F}^{ij}+
K_{ij}\tilde{K}^{ij})$. From the definition (3.1) and
the action (2.7) we obtain

\begin{eqnarray}
\tilde{K}^{ij} &=&  \frac{\partial L}{\partial F_{ij}}
\nonumber
\\ &=&
\frac{\sqrt{-detG_{ij}}}{\sqrt{-det(G_{ij}+
{\cal F}_{ij})}}(-{\cal F}^{ij}+{\cal T}\tilde{{\cal
F}}^{ij}) +
\tilde{\Omega}^{(1)ij},
\end{eqnarray}
where we have used the determinant formula for the
four-by-four matrix:
\begin{eqnarray}
det(G_{ij}+{\cal F}_{ij})=detG_{ij}(1+\frac{1}{2}
{\cal F}_{ij}{\cal F}^{ij} - {\cal T}^2),
{\cal T}\equiv \frac{1}{4}{\cal F}_{ij}\tilde{{\cal F}}^{ij}.
\end{eqnarray}
Taking the Hodge dual of (3.9), we find
\begin{eqnarray}
K_{ij}=-\frac{1}{2}\eta_{ijkl}\tilde{K}^{kl}
=\frac{\sqrt{-detG_{ij}}}{\sqrt{-det(G_{ij}+
{\cal F}_{ij})}}(\tilde{{\cal F}}_{ij}+{\cal T}{\cal
F}_{ij}) +\Omega_{ij}^{(1)}.
\end{eqnarray}
Then we obtain
\begin{eqnarray}
\frac{\lambda}{4}(F_{ij}\tilde{F}^{ij}+
K_{ij}\tilde{K}^{ij})
&=&\frac{\lambda}{4}(-2\Omega_{ij}^{(3)}
\tilde{F}^{ij} +2\Omega_{ij}^{(1)}
\tilde{K}^{ij} 
-\Omega_{ij}^{(1)}\tilde{\Omega}^{(1)ij}
-\Omega_{ij}^{(3)}\tilde{\Omega}^{(3)ij}) \nonumber \\
&=&\frac{\lambda}{4}(-4\Omega^{(3)} F
+4\Omega^{(1)}K
-2\Omega^{(1)}\Omega^{(1)}
-2\Omega^{(3)}\Omega^{(3)})
\end{eqnarray}
In the above and below we omit $\wedge$-symbol in products
of forms.

Next let us calculate $\delta_{\theta}L$. In the language of
differential forms,
\begin{eqnarray}
\delta_{\theta}L&=&\frac{\partial L}{\partial
{\cal F}}\delta\Omega^{(3)} + {\cal F}\delta\Omega^{(1)}
+\delta C_4 \nonumber \\
&=&\lambda[- K\Omega^{(1)}+(F+\Omega^{(3)})\Omega^{(3)}
+\frac{1}{2}(\Omega^{(1)}\Omega^{(1)}-
\Omega^{(3)}\Omega^{(3)})] \nonumber \\
&=&\lambda[-K\Omega^{(1)}+F\Omega^{(3)}
+\frac{1}{2}(\Omega^{(1)}\Omega^{(1)}+
\Omega^{(3)})\Omega^{(3)})].
\end{eqnarray}

It is clearly seen that the right-hand sides of (3-14) and (3-15)
exactly cancel with each other and the GZ-duality condition is 
indeed satisfied.
As we have proved the invariance of the action under the
infinitesimal $SO(2)$ duality transformation, the action is 
also invariant under the finite $SO(2)$ duality transformation.

 Before closing this section let us mension the extension
from the $SO(2)$ duality to the $SL(2,R)$ duality  by introducing 
a constant dilaton $\phi$ and an axion $\chi$ fields. We follow the
same line of arguments given in \cite{IIK1}. According to the
general method \cite{GR2}, let us define a new Lagrangian
$\hat{L}(G,F,\theta,\phi,\chi )$ from the D3-brane Lagrangian
$L(G, F ,\theta)$ which obey the $SO(2)$ duality ;
\begin{eqnarray}
\hat{L}(G,F,\theta,\phi,\chi )=L(G, e^{-\phi/2}F, \theta)
+\frac{1}{4}\chi F\tilde{F}.
\end{eqnarray}
Then if one define $\hat{F}= e^{-\phi/2}F$ and $\hat{K}$ by
taking the dual of $\frac{\partial L(G,\hat{F},\theta)}{\partial
\hat{F}}$, the background dependence is absorbed in the rescaled
variables $(\hat{K},\hat{F})$. These are related with the background
dependent $(K,F)$ by
%
\begin{gather}
\begin{pmatrix}
K \\
F \\
\end{pmatrix}
= V
\begin{pmatrix}
\hat{K} \\
\hat{F} \\
\end{pmatrix},
\end{gather}
\begin{gather}
V = e^{\frac{\phi}{2}}
\begin{pmatrix}
e^{-\phi} & \chi \\
0 & 1 \\
\end{pmatrix}.
\end{gather}
Here $V$ is a non-linear realization of $SL(2,R)/SO(2)$ transforming 
as
\begin{eqnarray}
V \longrightarrow V^{\prime}=\Lambda V O(\Lambda)^{-1}
\end{eqnarray}
Here $\Lambda$ is a global $SL(2,R)$ matrix
%
\begin{gather}
\Lambda = 
\begin{pmatrix}
 a & b \\
 c & d \\
\end{pmatrix}
\in SL(2,R), ad-bc=1,
\end{gather}
and $O(\Lambda)$ is an $SO(2)$ transformation
%
\begin{gather}
O(\Lambda)^{-1}=
\begin{pmatrix}
\cos\lambda & \sin\lambda  \\
-\sin\lambda & \cos\lambda  \\
\end{pmatrix}
\in SO(2).
\end{gather}
The condition that the form of $V$ (3.17) is unchanged under the
transformation (3.18) determins the $SO(2)$ rotation angle $\lambda$
and the transformation rule of the background fields $\phi$ and
$\chi$;
\begin{eqnarray}
\tan\lambda=\frac{ce^{-\phi}}{c\chi +d},
\end{eqnarray}
and
\begin{eqnarray}
\tau \rightarrow \tau^{\prime}=\frac{a\tau +b}{c\tau +d};
\tau \equiv \chi + ie^{-\phi}.
\end{eqnarray}

These results show that if the original Lagrangian $L(G,F,\theta)$ is
invariant under the $SO(2)$ duality transformation the extended
Lagarangian $\hat{L}(G,F,\theta,\phi,\chi )$
with a dilaton and an axion fields is invariant under the
$SL(2,R)$ duality transformation of $(K, F)$ and $\tau\equiv \chi+
ie^{-\phi}$ and $SO(2)$ rotation of $N=2$ spinor with rotation angle
$\lambda$ given by (3.21).

\section{Discussion}
\setcounter{equation}{0}
We have shown that the super D3-brane action on $AdS_5\times S^5$
background is exactly seif-dual, by proving that the action
satisfies the Gaillard and Zumino duality condition which is a
necessary and sufficient condition for the action to be self-dual. 
It should be stressed that our proof of the self-duality needs not
any gauge fixing choice for the $\kappa$-symmetry.

Although the super D3-brane action constructed by Metsaev and
Tseytlin for which we have proved the self-duality is based on the
explicit coset parametrization $G(X^{\hat{m}},\theta) =
g(X)e^{\theta Q}$, since the physical properties such as duality
relations among super p-branes should not depend on the choice of the
parametrization of manifolds, our result shoud be valid for any 
other parametrizations of coset manifold. 
In a recent paper \cite{PST} super p-brane theories based on a
different parametrizations of the coset manifolds are proposed. 
It would be interesting to confirm that
the super D3-brane action constructed based on this new coset 
parametrization is indeed self-dual.

 Now the self-duality of the super D3-brane action has been proved
for the flat and $AdS$ background. Therefore we expect that
the self-duality of super D3-brane and various duality relations
among super p-branes will be proved in generic background
geometries.


There are several effective actions of super D3-brane which
are proposed to be manifestly duality symmetric \cite{BBKN,CWT}.
It would be interesting to 
investigate possible relations with the duality symmetry  discussed
in this paper. 

 We may think about various $p$-branes on $AdS_5\times S^5$
background just as in the flat background. It would be interesting
to construct various super p-brane actions on $AdS_5\times S^5$ and
investigate  duality relations among them known on the flat
background in semiclassical approximation \cite{APPS}.

 It has been proposed that D-brane dualities are understandable as 
canonical transformation \cite{L,IIKK, IIK1}. It would be also
interesting to formulate dualities among super D-branes and M-branes
both on flat and AdS backgrounds as caninical transformations. 
 
\section*{Acknowledgments}
I would like to thank Y. Igarashi, K. Kamimura and
I. Oda for explaining their works.
This work is supported in part by the Grant-in-Aid for Research
from the Ministry of Education, Science and Culture. 


\end{document}